\documentclass[preprint,pre,aps,superscriptaddress,a4paper]{revtex4}
\usepackage[dvips]{graphicx}
\usepackage{epsfig,amsmath,amsfonts,amssymb,wasysym,color,hyperref}

\begin{document}

\title{Distribution of ions near a charged selective surface in critical binary solvents. }
\author{Alina Ciach}
\affiliation{\small Institute of Physical Chemistry, Polish Academy of
  Sciences,  Kasprzaka 44/52, PL-01-224 Warsaw, Poland.}
\author{Anna Macio\l ek}

\affiliation{\small Max-Planck-Institut f\"ur Metallforschung,
Heisenbergstr.~3, D-70569 Stuttgart, Germany.} 
\affiliation{\small Institut f\"ur Theoretische und
Angewandte Physik, Universit\"at Stuttgart, Pfaffenwaldring~57, D-70569 Stuttgart, Germany.}
\affiliation{\small Institute of Physical Chemistry, Polish Academy of
  Sciences, Kasprzaka 44/52, PL-01-224 Warsaw, Poland.}
\date{\today} 
\begin{abstract}

Near-critical binary mixtures containing ionic solutes  near a
 charged wall preferentially adsorbing one component of the solvent are studied. Within
the Landau-Ginzburg approach  extended to
 include  electrostatic interactions and the chemical preference of ions for
 one component of the solvent, we obtain a simple form for the  leading-order correction to the Debye-H{\"u}ckel theory result for the charge density profile. 
Our result shows that  critical
 adsorption influences significantly
 distribution of ions near the wall. This effect may have important implications for the screening of electrostatic interactions between charged surfaces immersed in  binary near-critical solvents.

\end{abstract}
\pacs{ 05.70.Jk,05.70.Np, 61.20.Qg, 68.35.Rh}
\maketitle

\section{Introduction}
\label{sec:intr}

A phenomenon of critical adsorption  occurs when a  fluid  is  brought to its bulk critical point in the presence of an attractive substrate or wall, for example, along the critical isochore. 
The wall causes a perturbation of the relevant order parameter profile to extend over a distance
 $\sim \xi_b$, the bulk correlation length, from the surface   \cite{binder:83:1,diehl:86:0}. Close to criticality,
 where
 $\xi_b\sim \mid( T-T_c)/T_c\mid^{-\nu} $ ($\nu$ is the critical exponent),
  the influence of the wall extends to macroscopic distances. As a result  the amount adsorbed
 (adsorption $\Gamma$)  diverges as  $\tau=(T-T_c)/T_c \to 0$.
Here we focus on the equivalent phenomenon that occurs for binary liquid mixtures near their consolute points. In these systems, generically there is a  preferential adsorption of one component of the mixture on the wall surfaces, resulting in the divergence of the relative adsorption near the critical point.

Critical adsorption was much studied both theoretically and experimentally. Of interest
has been the scaling prediction
 by Fisher and de Gennes \cite{FdG} that $\Gamma$ should have a power law dependence on  $\tau$, with
a universal 
exponent which does not depend on the details of the specific system. Also, this  phenomenon 
is  of significant practical importance, e.g., for the use of  supercritical fluids in micro- and nanofluidics \cite{supercritical} and as solvents in colloidal suspensions \cite{xia}.

In colloidal suspensions one usually encounters the  presence of ions. This  may be  due to the 
dissociation of a salt-free solvent, e.g., in binary solvents containing water, or due to dissolved salt. Ions influence the electrostatic interactions between the charged colloids leading to 
screening effects \cite{israelachvili}. If there is a chemical preference of ions for  one component of the solvent, then larger amount of ions may be dissolved in regions where the concentration of the preferred component of the solvent is larger. For this reason critical adsorption in binary solvents 
may affect the distribution of ions near charged surfaces. This effect in turn may influence the screening of electrostatic interactions between charged colloidal particles. 
Such a scenario is relevant  for an experimental system used recently  to investigate the effective forces acting on spherical particles  close to a substrate immersed in a near-critical water-lutidine mixture \cite{bechinger:08:01,GMHNHBD:09:01}. The aim of that study was to measure directly a critical Casimir force between a single spherical colloid and a flat surface. 
Polystyrene particles with high surface charge density were used as colloids, and silica glass treated chemically to achieve the desired adsorption properties, which also carried some surface charge, was used as  the substrate. Ions in solution were present due to dissociation in the salt-free  water-lutidine mixture.
In order to extract  the  critical Casimir force from the measured  potentials of interactions, the  distances  $z$ from the substrate
for which the electrostatic contribution was estimated to be  negligible have been considered. The estimation of the electrostatic interactions was based
on  measurements of the electrostatic potential  far from the critical point. 
However,  different solubilities of ions in water and lutidine and the enhancement of one of  components  of the near-critical mixture close to the surfaces might result in a change of the screening of the electrostatic interaction compared to the  case of a noncritical homogeneous medium. Unfortunately,
 possible effects of the  critical adsorption on the electrostatic contribution to the interaction potential could not be estimated due to the lack of any theoretical predictions for such effects.

Motivated by these issues in the  present paper we study   critical adsorption in a simple generic system, i.e., for a binary mixture solvent containing ionic solutes in the presence of  a single 
 charged planar wall that  preferentially adsorbs  one component of the solvent.

Recently, the properties of {\it bulk} near-critical binary mixtures with ions have been studied
theoretically using a  modified Poisson-Boltzmann theory by Onuki and Kitamura \cite{onuki:04:01}. The Landau-Ginzburg functional was used  to describe the
binary solvent. The presence of ions was accounted for  by  including  electrostatic contribution
and  the terms due to the  entropy of ions.  Preferential solvation effects were modeled by a linear coupling between the density of ions and the order parameter of  near-critical  binary mixture. 
Using this phenomenological functional in the mean-field approximation, the authors  found that the presence of ions shifts the critical point of demixing. This result is consistent with the 
 well known experimental observations  that adding  salt to the binary mixture solvent changes  the position of its phase separation curve,  such that the upper and lower critical points are shifted upwards~\cite{eckfeldt:43:0,hales:66:0} and downwards~\cite{balevicius:99:0}, respectively.

If the ion-containing mixtures are exposed to  external electric fields, e.g., due to  charged walls  or  charged macromolecules immersed in the solution, the phase separation temperature is also shifted
\cite{tsori:07:01}. In addition to the preferential solvation, a dielectric inhomogeneity is  important 
for this effect; the high permittivity solvent component is attracted to the charged surface thus
enhancing the  phase separation.
Recently, the effects of these, so called, dielectrophoretic forces and of the preferential solvation 
have been studied for ion-containing mixtures  confined between two-charged surfaces
in a single phase region  {\it away} from the coexistence  \cite{podgornik:09:01}.
The  approach used for a  bulk system in Ref.~\cite{onuki:04:01} have been modified
to  account for a spatial variation of the volume fractions and, 
consequently, of the dielectric permittivity of the binary mixture. In addition, a surface term describing the interactions between the charged solutes and  confining charged surfaces has been included.
Using a mean-field approximation the density profiles and osmotic pressure between charged interfaces 
have been calculated.

 In order  to investigate the influence of critical adsorption of ion-containing binary mixtures  
on the distribution of ions
near a single charged wall, we propose here an  approach that starts from the microscopic theory. Within such an approach the  Landau functional is derived rather than postulated, and it should describe correctly all the collective phenomena in the system that are consistent with the assumed interaction potentials. Hence, we can control all the assumptions on the fundamental level of interactions. 
In Sec.~\ref{subsec:constr} we introduce the lattice-gas model  of a four-component mixture describing
 two species of
 the solvent and the positive and 
negative  ions. The generic for this model, short-ranged interactions between  all the species
and between the species and the surface
are assumed. 
From the lattice model we derive the continuum Landau-Ginzburg functional which is then supplemented
by electrostatic bulk and surface contributions.  Contrary to previous approaches all parameters of our functional are expressed
in terms of microscopic interactions. Our Landau functional is similar to the functionals in Refs.~\cite{onuki:04:01,podgornik:09:01}, but more terms are present, and all the approximations are based on assumptions concerning types of interparticle interactions. 
 In Sec.~\ref{subsec:mf} we derive Euler-Lagrange  (EL) equations that allow to calculate the density and charge profiles on a mean-field level. Next, in Sec.~\ref{sec:so} we solve the linearized EL equations (Sec.~\ref{sec:solin}) and  calculate the leading-order correction
to them (Sec.~\ref{subsec:correct}) under the assumption of weak interactions with the surface, small amount of ions and weak surface charge.  A discussion of  our results is included.
The summary and outlook are given in Sec.~\ref{sec:sumout}.

\section{Derivation of the Landau Theory}
\label{sec:form}

\subsection{Construction of the lattice model}
\label{subsec:constr}

Let us consider a two-component mixture approaching  the upper critical point of the demixing transition from the one-phase side and investigate the effect of the presence of a small amount of positively and negatively charged ionic solutes. We are interested in such a system in contact with a charged wall, such that the total charge of the system neutralizes the charge at the wall. Let us first consider this system with the ions replaced by the corresponding neutral molecules. The electrostatic contribution to the free energy will be included in the next step. 

The four-component mixture close to the demixing transition can be conveniently studied in the framework of the lattice-gas model.
In the simplest lattice-gas model of a one-component system the space is divided into cubic cells of a volume comparable to the volume of a molecule.  The cells can be either occupied or unoccupied by a  center of a single molecule, i.e., multiple cell occupancy is forbidden. For a mixture of molecules of different sizes we consider a semi-microscopic description of cells occupied either by the larger molecule or by a cluster of smaller molecules, and we denote the volume of the lattice cell of a  simple cubic lattice (SC) by $v_0$. Here we limit ourselves to a mixture of molecules of similar sizes. The cell-occupancy operators are $\hat{o}_{i}({\bf x})=1$
if the cell ${\bf x}$ is occupied by the $i$-th component and $\hat{o}_{i}({\bf x})=0$
otherwise. 
The  two components of the solvent are denoted by $i=1, 2$, and two additional species that can be ionized to carry positive or negative charges are indicated by $i=3, 4$, respectively.
 We restrict ourselves  to thermodynamic conditions corresponding to the stability of a liquid, where total density fluctuations can be neglected. The single-state occupancy and close-packing lead to the constraint
\begin{equation}
\sum_{i=1}^{4}\hat{o}_{i}({\bf x})=1.\label{sso}
\end{equation}
 We consider an open system with the probability of the semi-microscopic state $\{\hat{o}_{i}({\bf x})\}$ given by
\begin{equation}
\label{p}
p[\{\hat{o}_{i}({\bf x})\}]=\frac{\exp(-\beta  H^{SR}[\{\hat{o}_{i}({\bf x})\}])\prod_{\bf x}\delta^{Kr}(\sum_{i=1}^{4}\hat{o}_{i}({\bf x})-1)}{\Xi},
\end{equation}
 where $\beta =(k_BT)^{-1}$, $k_B$ is the Boltzmann constant and $T$ the temperature. $\Xi$ is the partition function. We include the activity factors in the Boltzmann factor, by assuming the generalized Hamiltonian
\begin{equation}
 H^{SR}[\{\hat{o}_{i}({\bf x})\}]={\cal E}^{SR}[\{\hat{o}_{i}({\bf x})\}]-\mu_1\sum_{\bf x}\hat o_1({\bf x})-\mu_2\sum_{\bf x}\hat o_2({\bf x})-\mu_{3}\sum_{\bf x}(\hat o_3({\bf x})+\hat o_4({\bf x})),
\end{equation}
where $\mu_i$ is the chemical potential of the species $i$. We shall require charge neutrality for ionized species $3$ and $4$ in the bulk, and assume $\mu_3=\mu_4$.
For the short-range interaction energy we make the standard nearest-neighbor (NN) approximation for all the species
\begin{equation}
 {\cal E}^{SR}[\{\hat{o}_{i}({\bf x})\}]=-\sum_{\bf x}\sum_{n=1}^{3}\hat{o}_{i}({\bf x})J_{ij}\hat{o}_{j}({\bf x}+\hat{\bf e}_n),
\end{equation}
where $\hat{\bf e}_n$ is the unit lattice vector in the $n$-th direction and summation convention is assumed for $i,j=1,...,4$. The coupling constants are symmetric, $J_{ij}=J_{ji}$, and represent the sum of all short-range interactions.
The Hamiltonian of the lattice model with the electrostatic interactions disregarded takes the form
\begin{eqnarray}
\label{HSR}
 H^{SR}[\hat s,\hat\phi]= -\sum_{\langle{\bf x},{\bf x}'\rangle} \Bigg[\frac{1}{2}\Bigg(\hat s({\bf x})J_{ss}\hat s({\bf x}') + \hat\phi^2({\bf x})J_{\rho \rho}\hat \phi^2({\bf x}')+\hat \phi({\bf x})J_{\phi \phi}\hat \phi({\bf x}')\Bigg)
\\
\nonumber
 +\hat s({\bf x})J_{s\rho}\hat \phi^2({\bf x}')+ \hat\phi({\bf x})J_{\phi s}\hat s({\bf x}') +\hat\phi({\bf x})J_{\phi\rho}\hat \phi^2({\bf x}')\Bigg]
\\-\nonumber\sum_{{\bf x}\in V}\Bigg(\Delta\mu \hat s({\bf x})+
\Delta\mu_c \hat\phi^2({\bf x})\Bigg)
-\sum_{{\bf x}\in\partial V}\Bigg(h_s\hat s({\bf x})+h_{\rho}\hat\phi^2({\bf x})\Bigg)+const,
\end{eqnarray}
where $\sum_{\langle{\bf x},{\bf x}'\rangle}$ is the summation over all  pairs of nearest-neighbors  belonging to the system. The system volume is denoted by $V$,  $\partial V$ denotes the boundary layer, and the new  operators are 
\begin{eqnarray}
\label{eq:operators}
 \hat s=\hat o_1-\hat o_2=-1,0,1\\ \nonumber
\hat\phi=\hat o_3-\hat o_4=-1,0,1
\\ \nonumber
\hat \phi^2=\hat o_3+\hat o_4=0,1.
\end{eqnarray}
From  close packing (\ref{sso}) it follows that $\hat o_1+\hat o_2=\hat s^2=1-\hat \phi^2$.
The new coupling constants in (\ref{HSR}) are linear combinations of the coupling constants $J_{ij}$, and 
$\Delta\mu$ and $\Delta \mu_c$ are linear combinations of $\mu_i$. 

We further assume that the species $3$ and $4$ are similar, i.e., interact in a similar way with the solvent and with each other, such that we can assume
\begin{eqnarray}
\label{symmetries}
 J_{33}+J_{44}\approx 2J_{34}\\
\nonumber
J_{13}\approx J_{14}\\
\nonumber
J_{23}\approx J_{24}.
\end{eqnarray}
With this assumption 
the Hamiltonian  takes the simpler form
\begin{eqnarray}
\label{HSR1}
 {\cal H}^{SR}[\hat s,\hat\phi^2]=-\sum_{{\bf x}\in V}\sum_{n=1}^3\Bigg[\hat s({\bf x})J_{ss}\hat s({\bf x}+{\bf e}_n) + \hat\phi^2({\bf x})J_{\rho \rho}\hat\phi^2 ({\bf x}+{\bf e}_n)+\\
\nonumber
 \hat s({\bf x})J_{s\rho}\hat \phi^2({\bf x}+{\bf e}_n)+ \hat \phi^2({\bf x})J_{s\rho}\hat s({\bf x}+{\bf e}_n)\Bigg]
-\sum_{{\bf x}\in V}\Bigg(\Delta\mu \hat s({\bf x})+
\Delta\mu_c \hat \phi^2({\bf x})\Bigg)\\
\nonumber
-\sum_{{\bf x}\in\partial V}\Bigg(h_s\hat s({\bf x})+h_{\rho}\hat\phi^2({\bf x})\Bigg).
\end{eqnarray}
Note that the SR contribution to the energy depends  only on $\hat\phi^2$  when the assumptions (\ref{symmetries}) are valid; otherwise the full expression (\ref{HSR}) has to be considered. 
We assume $J_{ss}\gg J_{\rho\rho}$, so that the system can phase separate into two phases, one of them rich in the first- and the other one rich in the second component of the solvent, rather than into a phase rich in the $3$-rd component and  a phase rich in the $4$-th  component of the mixture.
The coupling $J_{s\rho}>0$ signals that ions preferentially dissolve in the first component of the solvent.

When the species $3$ and $4$ are  charged, then Coulomb interactions between them are present  in addition to the short range interactions. In the truly microscopic model one should take into account polarizability of the solvent, and the problem becomes very difficult. In the semi-microscopic  description the effect of polarizability of the solvent is included only through the dielectric permittivity. The  dielectric permittivity depends on the composition and, in the case of spatial inhomogeneities of the composition, the  dielectric permittivity is a functional of the solvent densities and depends on ${\bf x}$. The  electrostatic energy-density $e_e$ in this case  has the form \cite{jackson,onuki:04:01}
\begin{equation}
\label{ee}
 e_e[s({\bf x}),\rho_c({\bf x}),\hat\phi({\bf x}),\psi({\bf x})]=[-\frac{\epsilon({\bf x})}{8\pi}(\nabla \psi)^2+e\hat\phi({\bf x})\psi({\bf x})]
\end{equation}
where 
\begin{equation}
s({\bf x})=\langle\hat s ({\bf x})\rangle, \hskip1cm\rho_c({\bf x})=\langle\hat\phi^2({\bf x})\rangle
\end{equation}
are the mean values of  the microscopic operators, $\nabla$ is the  gradient and $\psi$ is the electrostatic potential that in equilibrium corresponds to the minimum of the energy (\ref{ee}), i.e., to generalization of the Poisson equation, considered for example in Ref.\cite{onuki:04:01}.
Close to the demixing  point of the binary mixture the  densities vary on the length scale of the bulk correlation length $\xi_b\gg v_0^{1/3}$, and their deviations from the average value are small. It is then reasonable to assume that $\epsilon({\bf x})$ is a linear function of the average densities of the two components of the solvent
\begin{equation}
\label{ep}
 \epsilon({\bf x})=\epsilon_1\rho_1({\bf x})+\epsilon_2\rho_2({\bf x})
\end{equation}
where $\rho_1({\bf x})=\langle\hat o_1({\bf x})\rangle=(1/2)(1-\rho_c({\bf x})+s({\bf x}))$ and $\rho_2({\bf x})=\langle\hat o_2({\bf x})\rangle=(1/2)(1-\rho_c({\bf x})-s({\bf x})) $.  $\epsilon_1$ and $\epsilon_2$ are the dielectric constants of the pure species 1 and 2 respectively. When the electrostatic energy has the above form, the average densities have to be determined selfconsistently. 

If one  assumes, however,  that  $\epsilon({\bf x})$ can be approximated by its average value
\begin{equation}
\label{epshomo}
 \epsilon({\bf x})\approx\bar\epsilon,
\end{equation}
then  the electrostatic energy (\ref{ee}) has the same form as in vacuum (with  the modified permittivity) and can be written as a sum of interaction energies for all pairs of point charges,
\begin{equation}
\label{Hch}
 {\cal E}^{C}[\{\hat{o}_{i}({\bf x})\}]=\frac{1}{2}\sum_{\bf x}\sum_{{\bf x'} \ne {\bf x} }\sum_{i,j=3,4}e_i\hat o_i({\bf x})V_c(|{\bf x}-{\bf x'}|)e_j\hat o_j({\bf x}')
\end{equation}
where $e_3$ and $e_4$ are the charges, and $V_c$ is the electrostatic interaction potential., that has the form different for different lattices and in the continuum space. The forms of $V_c$ for simple cubic, body centered, or face centered lattices in the Fourier representation can be found, for example, in Ref.\cite{ciach:04:0,ciach:05:1}. If we assume for $V_c$ the continuum-space form and restrict  the positions of ions  to the lattice sites, then the electrostatic energy takes the familiar form of the sum of Coulomb interaction potential for all the pairs of ions
\begin{equation}
\label{Hch1}
 {\cal E}^{C}[\{\hat\phi({\bf x})\}]=
\frac{e^2}{2}\sum_{\bf x}\sum_{{\bf x'}\ne {\bf x}}\frac{\hat \phi({\bf x})\hat \phi({\bf x}')}{\bar\epsilon|{\bf x}-{\bf x'}|}.
\end{equation}
In the above expression we limit ourselves to monovalent ions $e_3=-e_4=e$, with $e$ denoting the elementary charge.
   With the above form of the electrostatic energy we obtain a well-defined semi-microscopic model with the probability distribution
\begin{equation}
\label{p2}
p[\{\hat s,\hat \phi \}]=\frac{\exp(-\beta {\cal H}[\{\hat s,\hat \phi\}])\prod_{\bf x}\delta^{Kr}(\hat s^2+\hat\phi^2-1)}{\Xi},
\end{equation}
where ${\cal H}[\{\hat s,\hat \phi \}]= {\cal H}^{SR}[\{\hat s,\hat \phi^2 \}]+{\cal E}^{C}[\{\hat \phi \}]$. The grand potential in the above statistical-mechanical model is given by
\begin{equation}
 \Omega=-k_BT\ln\Xi.
\end{equation}
The advantage of such kind of modeling with the assumption (\ref{epshomo}) is the possibility for investigating the effect of fluctuations of the composition (for example by means of computer simulations), and the disadvantage is neglecting the coupling between $\epsilon$ (and hence the electrostatic energy of states $\{\hat o_i\}$)  and the composition fluctuations. Which effect, the fluctuations of the composition and their coupling to the density of ions or the spatial variation of $\epsilon$, plays the dominant role in determining the charge distribution near the charged wall  in a  system exhibiting critical fluctuations of concentration, remains an open question. 

\subsection{Mean-field approximation and continuous Landau-type model}
\label{subsec:mf}

 In the  mean-field (MF) approximation the grand potential 
is assumed to correspond to the minimum of the functional:
\begin{equation}
\label{OmMF}
 \Omega^{MF}[s,\rho_c,\phi,\psi]={\cal H}^{SR}[s,\rho_c]+\sum_{{\bf x}\in V}\left(e_e[s,\rho_c,\phi,\psi]-k_BTW[s,\rho_c,\phi]\right)
\end{equation}
where the electrostatic energy density $e_e[s,\rho_c,\phi,\psi]$ is given by (\ref{ee}), and $k_BW[s,\rho_c,\phi]$ is the entropy density. In this work we  focus on the semi-infinite system, and assume the lattice-gas form (ideal mixing entropy) 
\begin{eqnarray}
\label{W}
 W(z)=\frac{1-\rho_c(z)+s(z)}{2}\ln\Bigg(\frac{1-\rho_c(z)+s(z)}{2}
\Bigg)&+&\frac{1-\rho_c(z)-s(z)}{2}\ln\Bigg(\frac{1-\rho_c(z)-s(z)}{2}
\Bigg)    \nonumber \\
+\frac{\rho_c(z)+\phi(z)}{2}\ln\Bigg(\frac{\rho_c(z)+\phi(z)}{2}\Bigg)&+&\frac{\rho_c(z)-\phi(z)}{2}\ln\Bigg(\frac{\rho_c(z)-\phi(z)}{2}\Bigg),
\end{eqnarray} 
where we take into account the dependence of the fields on the distance $z$ from the planar wall.

We are interested in thermodynamic conditions corresponding to  stability of the uniform fluid close to the demixing transition. The equilibrium values of $s$ and $\rho_c$ in the bulk are the uniform solutions of the 
Euler-Lagrange  (EL) equations obtained from the minimization of the bulk part of the  functional (\ref{OmMF});
they are denoted by $\bar s$ and $\bar \rho_c$.
From now on we focus on the deviations from the equilibrium fields
\begin{eqnarray}
\label{fields}
 \vartheta(z)=s(z)-\bar s
\\
\eta(z)=\rho_c(z)-\bar\rho_c.
\end{eqnarray}

On the MF level it is possible to take into account the dependence of the dielectric constant on the composition and we assume that $\epsilon$ given by (\ref{ep}) depends on the distance $z$ from the planar wall according to
\begin{equation}
\label{ep1}
 \epsilon(z)=\bar\epsilon +\delta\epsilon(z),
\end{equation}
where 
\begin{equation}
\label{ep2}
 \delta\epsilon({\bf x})=\epsilon_{\vartheta}\vartheta({\bf x})-\epsilon_{\eta}\eta({\bf x}),
\end{equation}
and we introduced the notation
\begin{equation}
 \epsilon_{\vartheta}=\frac{(\epsilon_1-\epsilon_2)}{2}
\end{equation}
\begin{equation}
  \epsilon_{\eta}=\frac{(\epsilon_1+\epsilon_2)}{2}.
\end{equation}
For very small fields $\vartheta$ and $\eta$, the second term in Eq.(\ref{ep1}) is negligible compared to the first one. We shall keep this term in general formulas, valid for arbitrary deviations from bulk equilibrium densities.

The continuum-space  Landau-Ginzburg (LG) functional for the fields
$\vartheta(z)$ and $\eta(z)$ is defined as 
\begin{equation}
A {\cal L}[\vartheta,\eta,\phi,\psi]=\Omega^{MF}[\bar s+\vartheta,\bar\rho_c+\eta,\phi,\psi]-\Omega^{MF}[\bar s,\bar\rho_c,0,0],
\end{equation}
where $A$ is the area of the confining surface
in the semi-infinite geometry. $ {\cal L}[\vartheta,\eta,\phi,\psi]$ can be derived from the lattice model in a standard way. Here we follow the method desribed in detail in  Ref.\cite{ciach:04:2}. In short, in the first step EL equations on the lattice are derived, which  contain  contributions of the form $\nabla^2 f= f(z+1)+f(z-1)-2f(z)$ for $f=s,\rho_c$. For the boundary layer the effect of the missing neighbors is taken into account, and in this way surface EL equations are obtained. At the boundary layer both the bulk and the surface  EL equations must be satisfied, and the difference of the two equations gives the corresponding boundary condition. From the continuous version of the EL equations ($\nabla^2 f \rightarrow d^2 f/dz^2$) we derive the corresponding functional with the appropriate surface term that leads to the boundary conditions. We require that the boundary conditions contain lower-order derivatives than the bulk equations. The functional obtained in this way  has the form
\begin{eqnarray}
\label{functional}
 {\cal L}&=&\int_0^{\infty}dz\Bigg\{\frac{1}{2}{\mathbf v}^T(z){\bf C}^0\mathbf{v}(z)+\frac{1}{2}{\mathbf v'}^T(z){\bf J}{\mathbf  v'}(z)+e_e+k_BTP
\Bigg\} \\
\nonumber &+&\frac{{\mathbf v}^T(0){\mathbf J}{\mathbf v}(0)}{2}-{\mathbf h}{\mathbf v}(0)+e\sigma\psi(0),
\end{eqnarray}
where  boldface capital letters denote matrices. The transpose of the columnar vectors ${\bf v}$ and  ${\bf v}'$ are ${\mathbf v}^T(z)=(\vartheta(z),\eta(z))$ and ${\mathbf v'}^T(z)=(d\vartheta(z)/dz,d\eta(z)/dz)$ respectively. The elements of the  matrix  ${\bf C}^0=(C^0_{ij})_{i,j=1,2}$, where  indices $1$ and $2$  correspond to $s$ and $\rho$, respectively, are given by 
\begin{eqnarray}
\label{C0}
 C^0_{ss}= k_BT\frac{1-\bar\rho_c}{(1-\bar\rho_c)^2-\bar s^2}-6J_{ss}\\
C^0_{\rho\rho}= k_BT\Bigg(\frac{1-\bar\rho_c}{(1-\bar\rho_c)^2-\bar s^2}+\frac{1}{\bar\rho_c}\Bigg)-6J_{\rho\rho}\\
C⁰^0_{s\rho}=C^0_{\rho s}=k_BT\frac{\bar s}{(1-\bar\rho_c)^2-\bar s^2}-6J_{\rho s}.
 \end{eqnarray}
The second matrix in (\ref{functional}) is ${\bf J}:=(J_{ij})_{i,j=1,2}$, again with indices $1$ and $2$  corresponding to $s$ and $\rho$, respectively. 
The  electrostatic energy-density $e_e$ in the  case of  a fluctuating dielectric constant has the form (\ref{ee}).
 The linear surface fields ${\mathbf h}=(\bar h_s,\bar h_{\rho})$   are 
\begin{equation}
 \bar h_s=h_s-J_{ss}\bar s-J_{s\rho}\bar \rho_c, \qquad \bar h_{\rho}=h_{\rho}-J_{\rho \rho}\bar\rho_c-J_{\rho s}\bar s,
\end{equation}
 and $P$ is obtained from the expansion of $W$ about  $s=\bar s$ and $\rho_c=\bar\rho_c$  with the linear and quadratic parts in the fields $\vartheta$ and $\eta$ subtracted.
The length unit is the lattice constant $v_0^{1/3}$, comparable with the molecular diameter. 
The last term describes the electrostatic interaction with the uniformly charged
 surface possessing the surface charge density $e\sigma$.

\subsection{Spinodal surface and bulk correlation functions}
\label{sec:spinsurf}

In order to find the bulk correlation functions in the Gaussian approximation as well as the spinodal surface and the critical line of this model in MF, let us consider the Gaussian part  of the LG functional in the bulk. In the Fourier representation we have, 
\begin{eqnarray}
\label{LG}
{\cal L}_G = \int \frac{d{\bf k}}{(2\pi)^3}\frac{1}{2}\tilde{\mathbf{v}}^{\ast}({\mathbf k})\tilde{{\bf C}}(k)\tilde {\mathbf{v}}(\mathbf{k})
\end{eqnarray}
where $\tilde{\mathbf{v}}^{\ast}({\mathbf k})=\tilde{\mathbf{v}}^{T}(-\mathbf{k})$,  and 
\begin{equation}
 \tilde {\bf C}(k)={\bf C}^0+k^2{\bf J}={\bf J}({\bf J}^{-1}{\bf C}^0+k^2{\bf I}).
\end{equation}
The correlation functions for the fluctuations of the solvent composition $\vartheta$ and the density of ions $\eta$ in the Fourier representation are given by the matrix $\tilde {\bf G}=\tilde {\bf C}^{-1}$. Each component of this matrix is inversely proportional to $\det ({\bf J}^{-1}{\bf C}^0+k^2{\bf I}) =(\lambda_1^2+k^2)(\lambda_2^2+k^2)$, where $\lambda_1^2$ and $\lambda_2^2$ are the two eigenvalues of the matrix 
\begin{equation}
\label{M}
{\bf M}={\bf J}^{-1}{\bf C}^0
\end{equation}
and are given by
\begin{eqnarray}
\label{eq:lambdy}
\lambda_{1,2}^2= \frac{Tr{\bf M}\mp\sqrt{\left(Tr{\bf M}\right)^2-4 \det {\bf M}}}{2}.
\end{eqnarray}
The correlations are thus given by a linear combination of the two terms, 
$ (\lambda_1^2+k^2)^{-1}$ and  $ (\lambda_2^2+k^2)^{-1}$. 
The asymptotic decay of correlations in real space is described  by the  inverse correlation length $\xi^{-1}_b=\min(\lambda_1,\lambda_2)$.

The uniform  phase is stable in the range of parameters $T, {\bar s}$ and ${\bar\rho_c}$ such that
 $\det \tilde {\bf C}(k)>0$ for all $k<k_{max}$, where $k_{max}$ is an  upper cutoff
on wave
numbers $\left|{\mathbf k}\right| \le k_{max}=\pi / a$. $a$ is usually identified with some
appropriate microscopic length, e.g., the lattice spacing or the molecular diameter.  Because $\det \tilde {\bf C}(k)=J\det \left( {\bf M}+k^2\mathbf{I}\right)=J(\lambda_1^2+k^2)(\lambda_2^2+k^2)$ 
   where  
\begin{eqnarray}
\label{J}
J=\det{\bf J}=J_{ss}J_{\rho\rho}-J^2_{s\rho},
\end{eqnarray}
 for the instability  analysis we have to distinguish  cases of positive and negative  $J$.
 
 If $J>0$, from the condition $\det  \tilde {\bf C}(0)\ge 0 $ (one-phase stability) we obtain after some algebra that $Tr{\bf M}\ge 0$ 
 and  $\lambda_1^2, \lambda_2^2>0$. 
In that case the instability of the uniform phase occurs at $k=0$ when  
\begin{eqnarray}
\label{instability}
 \det \tilde {\bf C}(0)=J\det {\bf M}=J\lambda_1^2\lambda_2^2= 0 \qquad {\rm or}\qquad  \tilde C_{ss}=0
\end{eqnarray}
The condition $\tilde C_{ss}=0$  determines the instability for the case of the vanishing field $\eta$ (the same concentration of ions in the phase-separated solvent).
From (\ref{eq:lambdy}) it follows that  $\lambda_1^2\to 0^{+}$ so that the asymptotic inverse decay length is $1/\xi_b=\lambda_1$.

For $J<0$, on the other hand, the stability condition of the uniform phase,  $\det \tilde {\bf C}(k)>0$, 
requires $\det \left({\bf M}+k^2\mathbf{I}\right) =(\lambda_1^2+k^2)(\lambda_2^2+k^2)<0$ for all $k<k_{max}$. In particular,  the uniform phase is stable against macroscopic phase separation as long as $\lambda_1^2\lambda_2^2<0$. The model can describe phase separation into two uniform phases only when the boundary of stability of the uniform phase is associated with $k=0$, i.e., when upon decreasing temperature
$ (\lambda_1^2+k^2)(\lambda_2^2+k^2)$ changes sign first at  $k=0$. In our case this means that  $\lambda_1^2\lambda_2^2>(\lambda_1^2+k_{max}^2)(\lambda_2^2+k_{max}^2)$;  the necessary condition for the above is $\lambda_1^2+\lambda_2^2<0$. This condition can only be satisfied when
 $Tr{\bf M}<0$, since in this case  $\lambda_1^2<0$ and $\lambda_2^2\to 0^{+}$; hence,  $1/\xi_b=\lambda_2$. 
Our model permits an instability at  $0<k\le k_{max}$, which will lead to the modulated phase,
 for a set of parameters  satisfying $J<0$ and $Tr{\bf M}>0$, since in this case  $\lambda_1^2\to 0^{-}$ and $\lambda_2^2>0$. We will not consider such parameters in the current paper.

For a given composition, 
the  spinodal  surface $T_s({\bar s},{\bar\rho_c}) $
is determined by  the higher temperature satisfying the condition (\ref{instability}).
The explicit relation between the temperature and the composition of the mixture that satisfies the first condition in  Eq.~(\ref{instability}),  reads
\begin{eqnarray}
\label{spin}
(k_BT_s)^2 &+&
k_BT_s\left[12{J}_{s\rho}{\bar s}{\bar \rho_c}-6J_{ss}(1-{\bar \rho_c}-{\bar s}^2)-6J_{\rho\rho}{\bar\rho_c}(1-{\bar\rho_c})\right] \\ \nonumber
& + & 36 J{\bar\rho_c}((1-{\bar\rho_c})^2-{\bar s}^2) =0
\end{eqnarray}
For $\bar\rho_c=0$ we have simply $k_BT_c=6J_{ss}$ and ${\bar s}_c=0$ in this model.  For a very small amount of ions the shift of the critical temperature can be estimated from  the derivative of $T_s$ with respect to $\bar \rho_c$ at $\bar s=\bar \rho_c=0$. The derivative at that point is
\begin{equation}
 \frac{d T_s}{d \bar \rho_c}=\frac{6(J_{s\rho}^2-J_{ss}^2)}{J_{ss}},
\end{equation}
while the derivative at $\bar s=\bar \rho_c=0$ of $k_BT$ satisfying $\tilde C_{ss}(0)=0$ is $-6J_{ss}$. Hence, Eq. (\ref{spin}) describes the actual instability at least for small amount of the solute. 
The critical temperature increases upon addition of ions when $J_{s\rho}-J_{ss}>0$. Experimentally addition of some salt leads to the increase of  the temperature of the upper critical point \cite{eckfeldt:43:0,hales:66:0}. Thus, if we require that this model describes the experimental situation on the MF level, we should assume that $J_{s\rho}-J_{ss}>0$. This condition, together with the earlier assumption $J_{ss}>J_{\rho\rho}$, leads to $J< J_{ss}^2-J_{s\rho}^2<0$. On the other hand, for $J>0$  the critical point decreases upon addition of  salt, which is not consistent with the experimental observation. However, the present study is limited to MF approximation, and the effect of fluctuations on the critical point should be taken into account to draw definite conclusions. We shall consider both cases, $J>0$ and $J<0$, bearing in mind that in the latter case the parameters must satisfy the condition $Tr {\bf M}<0$.

 \subsection{Euler-Lagrange equations  in semi-infinite system}
 \label{subsec:sis}

Our goal is to determine  the shape of  the ion number and charge density profiles near the charged wall, therefore we need to consider the Euler-Lagrange (EL) equations corresponding to the minimum of  the functional (\ref{functional}).

From $\delta {\cal L}/\delta \phi(z)=0$ we obtain
\begin{eqnarray}
\label{pii}
 \phi(z)=-\rho_c(z)g(z),
\end{eqnarray}
where
\begin{eqnarray}
 g(z)=\tanh(\beta e \psi(z)/2).
\end{eqnarray}
Note that $-1<g(z)<1$.  From the above  we can obtain $\psi$ as a Taylor series in $\phi$ and $\eta$, of odd orders in $\phi$. The charge density can be eliminated by using (\ref{pii}), and after some algebra we 
obtain three  EL equations for $\vartheta$, $\eta$ and $g$:
\begin{eqnarray}
\label{EL}
 J(\vartheta^{''}+6\vartheta)
=\frac{k_BT}{2}\Bigg[(J_{\rho\rho}+J_{\rho s})\ln(1-\rho_c+s)+(J_{\rho s}-J_{\rho\rho})\ln(1-\rho_c-s) \\ \nonumber
-2J_{\rho s}\ln(\rho_c)-J_{\rho s}\ln(1-g^2)-J_{\rho\rho}\Delta \bar\mu +J_{s\rho}\Delta \bar\mu_c
\Bigg]\nonumber
\end{eqnarray}
\begin{eqnarray}
\label{EL1}
 J(\eta^{''}+6\eta)
=-\frac{k_BT}{2}\Bigg[(J_{\rho s}+J_{ss})\ln(1-\rho_c+s)+(J_{ss}-J_{\rho s})\ln(1-\rho_c-s)\\
\nonumber
-2J_{ss}\ln(\rho_c)-J_{ss}\ln(1-g^2)+J_{s\rho}\Delta \bar\mu -J_{ss}\Delta \bar\mu_c \Bigg]
\end{eqnarray}
\begin{eqnarray}
\label{EL2}
 -g^{''}\Big(\bar\epsilon+\epsilon_{\vartheta}\vartheta-\epsilon_{\eta}\eta\Big)  +g^{'}g\frac{e}{k_BT}\Big[(\bar\epsilon+\epsilon_{\vartheta}\vartheta^{'}-\epsilon_{\eta}\eta^{'}\Big] -2\pi e\left(\frac{e}{k_BT}\right)^2(1-g^2)g^2\rho_c=0,
\end{eqnarray}
where  
\begin{eqnarray}
 \Delta \bar \mu = \Delta \mu +6J_{ss}\bar s + 6J_{\rho s}\bar \rho_c=-\frac{k_BT}{2}\ln\Big[\frac{1-\bar\rho_c+\bar s}{1-\bar\rho_c-\bar s}\Big]
\end{eqnarray}
and   
\begin{eqnarray}
 \Delta \bar \mu_c = \Delta \mu_c +6J_{\rho\rho}\bar \rho_c + 6J_{\rho s}\bar s=-\frac{k_BT}{2}\ln\Big[\frac{\bar\rho_c^2}{(1-\bar\rho_c)^2-{\bar s}^2}\Big].
\end{eqnarray}
In the above equations we used the explicit lattice-gas  expressions for the entropy (\ref{W}).
Equations for $\Delta \bar \mu$ and   $\Delta \bar \mu_c$ follow from the bulk EL equations.

The boundary conditions of the EL equations for the functional (\ref{functional}) are
\begin{equation}
\label{en_el}
\frac{{\bar\epsilon}+\epsilon_{\vartheta}\vartheta(0)- \epsilon_{\eta}\eta(0)}{4\pi}\nabla\psi(z)\vert_{z=0} =-e\sigma
\end{equation}
and 
\begin{equation}
\label{bc}
 \vartheta'-\vartheta=H_{\vartheta}, \qquad \eta'-\eta=H_{\eta},
\end{equation}
where
\begin{equation}
\label{bc1}
 H_{\vartheta}=\frac{\bar h_{\rho}J_{\rho s}-\bar h_sJ_{\rho\rho}}{J}=\bar s+\frac{h_{\rho}J_{\rho s}-h_sJ_{\rho\rho}}{J}, \qquad H_{\eta}=\frac{\bar h_sJ_{\rho s}-\bar h_{\rho}J_{ss}}{J}=\bar\rho_c +\frac{h_sJ_{\rho s}-h_{\rho}J_{ss}}{J}.
\end{equation}
$J$ is defined in (\ref{J}).

When $\epsilon=\bar\epsilon$, the third EL simplifies, 
\begin{eqnarray}
 g^{''}(z)-2g^{'}g(z)(e/k_BT)+(2\pi e/\bar \epsilon)(e/k_BT)^2\rho_c(z)g(z)^2(1-g(z)^2)=0.
\end{eqnarray}
Alternatively, $\psi$ could be eliminated, and three EL equations for the fields $\vartheta, \eta$ and $\phi$ would be obtained.
 
\section{Approximate solutions of the EL equations}
\label{sec:so}
 
 The nonlinear EL equations (\ref{EL})-(\ref{bc}) can be solved numerically. Since in the critical region and for weak surface fields and small $\sigma$ the magnitudes of the fields $\phi$, $\vartheta$ and $\eta$ are small, we can obtain approximate analytical solutions within the perturbation method. Analytical solutions can give more general insight, but in the future studies should be supplemented with numerical solutions for particular choices of model parameters.

We postulate that the solutions of the EL equations can be written in the form
\begin{eqnarray}
\label{expan}
f=\sum_{n=1}^Nf^{(n)}
\end{eqnarray}
where  $f=\vartheta,\eta,\phi,\psi$,  $f^{(n)}=O(\nu^n)$, and $\nu$ is a small parameter. 
\subsection{Solution of the linearized EL equations}
\label{sec:solin}
Let us first consider the linearized equations (\ref{EL})- (\ref{EL2}). For the linearized equations we shall simplify the notation, $\vartheta^{(1)}\equiv \vartheta, \eta^{(1)}\equiv \eta,\phi^{(1)}\equiv \phi$. 
For the electrostatic part we obtain the linearized equations of the well known form (the spatial 
dependence of permittivity leads to nonlinear contributions to the EL equations)

\begin{eqnarray}
 \frac{\bar\epsilon}{4\pi}\frac{d^2\psi(z)}{dz^2}+e\phi(z)=0
\end{eqnarray}
\begin{eqnarray}
 e\psi(z)+\frac{kT\phi(z)}{\bar\rho_c}=0
\end{eqnarray}
which together give 
\begin{eqnarray}
\label{eq:electr}
 \phi(z)^{''}=\kappa^2\phi(z),
\end{eqnarray}
where 
\begin{eqnarray}
 \kappa^2=\frac{4\pi e^2\bar\rho_c}{kT\bar\epsilon}.
\end{eqnarray}
 Note that $\kappa$ is the inverse Debye length in units of the molecular size.
Solution of Eq.~(\ref{eq:electr}) with the boundary condition Eq.~(\ref{en_el}) has the well known form
\begin{eqnarray}
\label{wellknown}
 \phi(z)=-\kappa \sigma e^{-\kappa z}.
\end{eqnarray}

The equations for ($\eta,\vartheta$) are formally  the same as in the Landau theory for a mixture near the demixing critical point, and 
can be written in the form
\begin{eqnarray}
\label{EL0}
\mathbf v''={\bf M} \mathbf v
\end{eqnarray}
where the matrix ${\bf M}=(M_{ij})$ with the indices $1$ and $2$  corresponding to $s$ and $\rho$ respectively, is defined in (\ref{M}), and the vector ${\bf v}$ is defined below Eq.(\ref{functional}).
In the semi-infinite systems in the one-phase  state  $\vartheta(z), \eta(z) \to 0$ for $z\to \infty$, and 
from the Ansatz:
\begin{eqnarray}
\label{eq:semi}
 \vartheta=t_1\exp(-\lambda_1 z)+t_2\exp(-\lambda_2 z)\\
\eta=n_1\exp(-\lambda_1 z)+n_2\exp(-\lambda_2 z)
\end{eqnarray}
we obtain
\begin{eqnarray}
\label{ratioam}
 \frac{t_i}{n_i}=\frac{\lambda^2_i-M_{22}}{M_{21}}.
\end{eqnarray}
 For $T\to T_c$ we have $1/\xi_b\to 0$ where  $1/\xi_b=\lambda_1$ or $1/\xi_b=\lambda_2$ for $J>0$ or $J<0$ respectively (see sec.\ref{sec:spinsurf}), and  the equations (\ref{eq:semi}) take the form
\begin{eqnarray}
\label{eq:limit}
 \vartheta(z)\simeq t_1\exp(-z/\xi_b )\\
\eta(z) \simeq n_1\exp(-z/\xi_b ).
\end{eqnarray}
From the boundary conditions and Eq.(\ref{ratioam}) we obtain in this case 
\begin{eqnarray}
\label{n1to}
 n_1\simeq_{T\to T_c} -\frac{(H_{\eta}M_{11}-H_{\vartheta}M_{21})}{Tr{\bf M}}+ ...
\end{eqnarray}
\begin{eqnarray}
\label{t1to}
 t_1\simeq_{T\to T_c} -n_1\frac{M_{22}}{M_{21}}+ ...
\end{eqnarray}

Note that the linearized equations for $\phi$ and ($\eta,\vartheta$) are decoupled. Thus, at the linear order there is no effect of the concentration profile on the charge distribution. Similar result was obtained recently in Ref.\cite{bier}. Note, however that we consider a simplified model of SR interactions, Eq.(\ref{HSR1}), where direct SR couplings between the charge and the concentration are disregarded. 

\subsection{Leading-order correction to the linearized EL equations}
\label{subsec:correct}

Beyond the linear order the coupling between the EL equations leads to  modifications of the charge and concentration profiles. Note that the solution of the nonlinear equations can be written in the form (\ref{expan}) when all amplitudes of the solutions of the linearized equations are  small and of the same order, $\kappa\sigma=O(\nu)$, $n_i=O(\nu)$ and $t_i=O(\nu)$.  This means a large Debye screening length and a small surface-charge density, as well as weak surface fields $H_{\eta},H_{\theta}$ (see Eqs. (\ref{n1to}) (\ref{t1to})). We shall limit ourselves to such conditions and calculate 
 the first correction $\vartheta^{(2)}$, $\eta^{(2)}$, $\phi^{(2)}$  to the solutions of the linearized equations. 

\begin{figure}
 \includegraphics[scale=0.38]{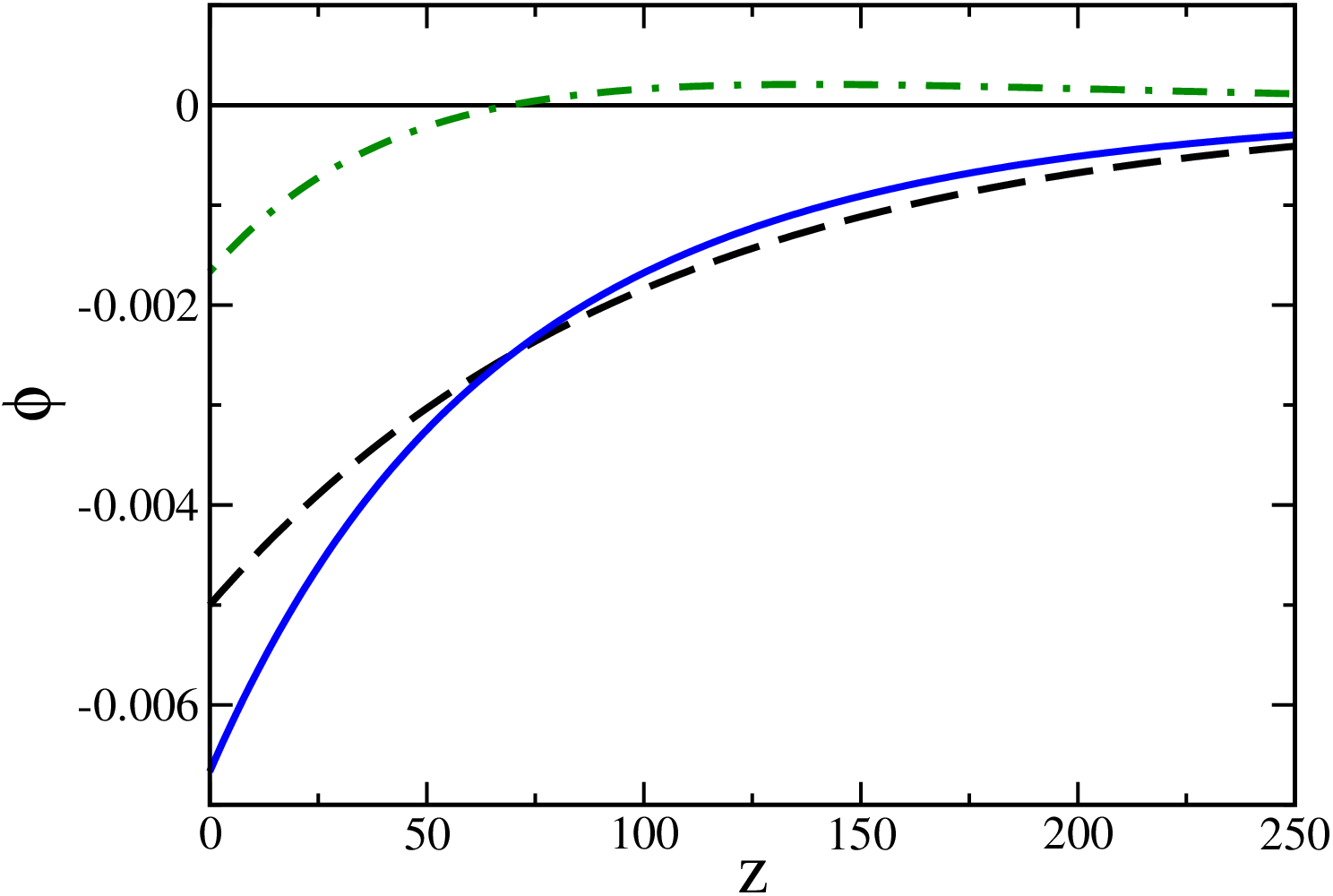}
\caption{ (Color online) Charge density profiles in the semi-infinite system. The dashed line is the solution of the linearized equation (\ref{wellknown}), $\phi^{(1)}(z)$, the dash-dotted line is the correction term (\ref{ra}), $\phi^{(2)}(z)$, and the  solid line is the sum of the two functions, $\phi^{(1)}(z)+\phi^{(2)}(z)$, for $\kappa=0.01$, $\sigma=0.5$, $\xi_b=100$, and $n_1=0.5$. The charge density is in $e/v_0$ units, where $e$ is the elementary charge, and $z$ is in $v_0^{1/3}$ units, where $v_0$ is the volume per molecule in the close-packed system and we assume that all molecules are of  similar size. 
}
\label{fig:1}
\end{figure}

\begin{figure}
\includegraphics[scale=0.38]{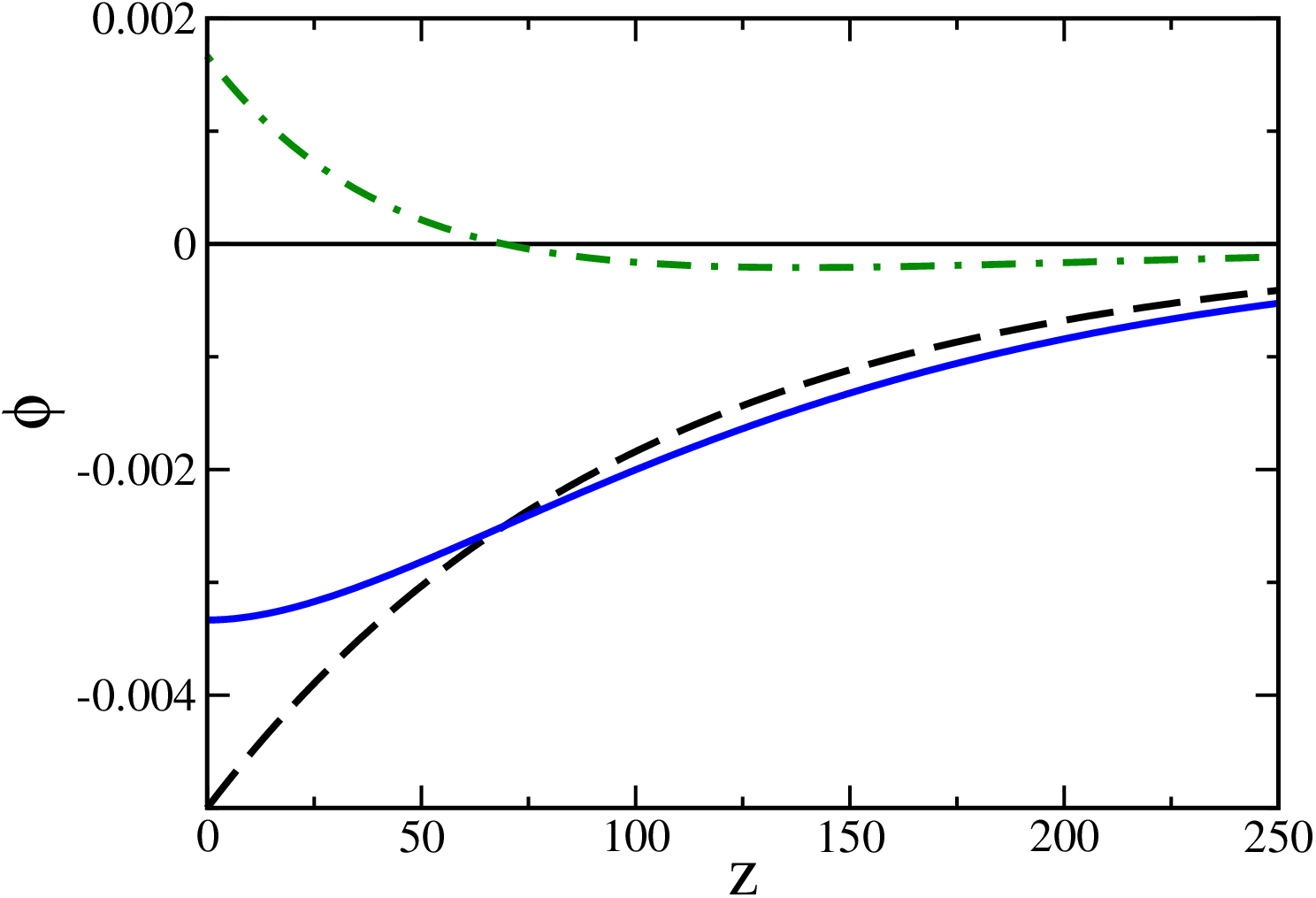}
\caption{ (Color online) Charge density profiles in the semi-infinite system. The dashed  line is the solution of the linearized equation (\ref{wellknown}),  $\phi^{(1)}(z)$, the dash-dotted line is the correction term (\ref{ra}),  $\phi^{(2)}(z)$, and the solid  line is the sum of the two functions  $\phi^{(2)}(z)$, for $\kappa=0.01$, $\sigma=0.5$, $\xi_b=100$, and $n_1=-0.5$. The charge density is in $e/v_0$ units where $e$ is the elementary charge, and $z$ is in $v_0^{1/3}$ units, where $v_0$ is the volume per molecule in the close-packed system and we assume that all molecules are of  similar size. 
}
\label{fig:2}
\end{figure}

\begin{figure}
\includegraphics[scale=0.38]{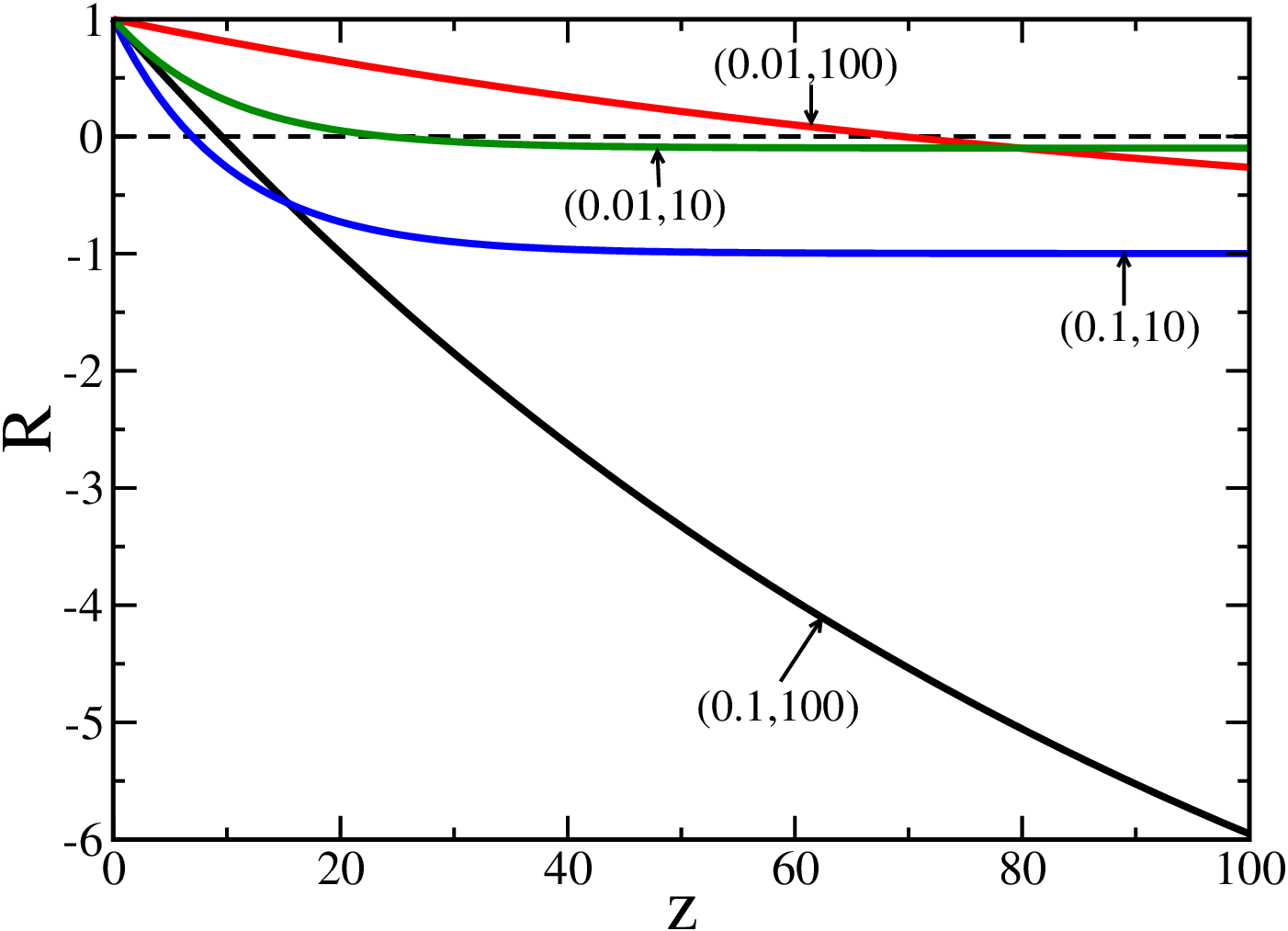}
\caption{ (Color online) The rescaled ratio $R$ (Eq.(\ref{R})) between the leading-order correction (\ref{ra}) and the Debye-H\"uckel form (\ref{wellknown}) of the charge density profile for several choices of
pairs $(\kappa,\xi_b)$ as indicated in the plot. $z$ is in $v_0^{1/3}$ units, where $v_0$ is the volume per molecule in the close-packed system and we assume that all molecules are of  similar size. 
}
\label{fig:3}
\end{figure}

For the electrostatic part 
we obtain
\begin{eqnarray}
\label{eq:pop}
 \psi^{(2)}= \frac{k_BT}{e\bar\rho_c}\Bigg[\frac{\phi\eta}{\bar\rho_c}-
\phi^{(2)}
\Bigg]
\end{eqnarray}
and
\begin{eqnarray}
\label{c2}
- \left(\phi^{(2)}\right)^{''}+\kappa^2\phi^{(2)}+{\cal N}_{\phi}(z)=0
\end{eqnarray}
where
\begin{eqnarray}
\label{eq:c2a}
 {\cal N}_{\phi}(z)=\frac{\kappa^2}{4\pi e}\Bigg[
(\epsilon_{\vartheta}\vartheta^{'}(z)-\epsilon_{\eta}\eta^{'}(z))\psi^{'}(z)+
(\epsilon_{\vartheta}\vartheta(z)-\epsilon_{\eta}\eta(z))\psi^{''}(z)\Bigg]+(\phi(z)\eta(z))^{''}
\end{eqnarray}
Note that at the second order in the perturbation expansion we need to calculate the function ${\cal N}_{\phi}$ to the quadratic order in $\nu$. The first term in Eq.(\ref{eq:c2a}) is of the order $O(\nu^2\kappa^2)$ (recall that we limit ourselves to boundary conditions such that the fields $\vartheta,\eta,\phi$ obtained from linearized equations are all of the order of $O(\nu)$). 
The perturbation expansion (\ref{expan}) is justified when $\kappa\sigma=O(\nu)$ (see (\ref{wellknown})), i.e., for small surface charge and/or  weak screening. When the additional condition, namely $\kappa^2\ll 1$ is satisfied (weak screening), the first term in (\ref{eq:c2a}) is negligible compared to the second one. In this case of weak screening the position-dependence of the dielectric constant does not play a dominant role for the leading-order correction to the solution of the linearized EL equations, and we may assume $\epsilon=\bar\epsilon$. 
From the forms of the fields $\vartheta,\eta,\phi$ obtained from linearized equations we obtain the explicit expression for $ {\cal N}_{\phi}(z)$, and
Eq.(\ref{c2}) simplifies to
\begin{eqnarray}
 - \left(\phi^{(2)}\right)^{''}+\kappa^2\phi^{(2)}-\kappa \sigma e^{-\kappa z}\Big[ (1/\xi_b+\kappa)^2 n_1e^{-z/\xi_b }\Big]=0.
\end{eqnarray}
The solution is
\begin{eqnarray}
\label{phi2}
 \phi^{(2)}(z)={\cal A}_0 e^{-\kappa z}+{\cal A}\xi_be^{- (\kappa+1/\xi_b)z},
\end{eqnarray}
where 
\begin{eqnarray}
\label{eq:A}
 {\cal A}=-\frac{\kappa \sigma n_1(\xi^{-1}_b+\kappa)^2}{(\xi^{-1}_b+2\kappa)}
\end{eqnarray}
and 
 ${\cal A}_0$ can be determined from the electroneutrality condition, 
\begin{eqnarray}
 \int_0^{\infty}dz(\phi^{(1)}(z)+\phi^{(2)}(z))+\sigma=0.
\end{eqnarray}
The above condition is satisfied by the solution (\ref{wellknown})  of the linearized equation, hence for the correction term we have
the condition 
 $\int_0^{\infty}dz\phi^{(2)}(z)=0$
that  yields 
\begin{eqnarray}
\label{eq:A0}
{\cal    A}_0=-\kappa \frac{{\cal A}\xi_b}{(\kappa+\xi^{-1}_b)}.
\end{eqnarray}
 In the critical region $\xi^{-1}_b\to 0$, hence ${\cal A}\xi_b\to \infty$. However, $ \phi^{(2)}(z)$ can be written in the following  form 
\begin{eqnarray}
\label{corterm}
 \phi^{(2)}(z)=\bar{\cal A}_0 e^{-\kappa z}+\frac{\kappa\sigma n_1(\xi^{-1}_b+\kappa)^2}{(\xi^{-1}_b+2\kappa)}e^{-\kappa z}\frac{(1-e^{-z/\xi_b })}{\xi^{-1}_b}
\end{eqnarray}
where 
\begin{eqnarray}
\label{eq:A0bar}
 \bar{\cal A}_0 = -\kappa\sigma\frac{ n_1(1+\xi_b\kappa)}{1+2\xi_b\kappa}.
\end{eqnarray}
Thus, $  \phi^{(2)}(z)$ is regular for  $\xi^{-1}_b\to 0$, but the second term in (\ref{corterm}) is non-monotonic as a function of $z$.
The ratio between the correction and the leading-order term has the form
\begin{eqnarray}
\label{ra}
 \frac{ \phi^{(2)}(z)}{  \phi^{(1)}(z)}\approx\frac{ n_1(1+\xi_b\kappa)}{1+2\xi_b\kappa} R(z),
\end{eqnarray}
where 
\begin{eqnarray}
\label{R}
 R(z)=
\Bigg[1 -(1+\xi_b\kappa)(1-e^{-z/\xi_b })\Bigg]
\end{eqnarray}
is independent of the amplitude $n_1$, i.e.  $\phi^{(2)}(z)/ \phi^{(1)}(z)\propto n_1$.
The coupling between the concentration and the charge fluctuations in the functional (\ref{functional}) leads to the non-monotonic correction to the linear solution of the EL equations, with the extremum at
\begin{eqnarray}
z_{extr}=2\xi_b\ln\Bigg(\frac{\kappa+\xi^{-1}_b}{\kappa}\Bigg) .
\end{eqnarray}
From Eq.~(\ref{ra})  it follows that the correction term changes its sign at $z_0=z_{extr}/2$, i.e., the critical adsorption  for $z<z_0$ leads to the charge density {\it larger (smaller)} than that predicted by the Debye-H\"uckel theory for $n_1>0 \quad (n_1<0)$, whereas for $z>z_0$ the charge density is {\it smaller (larger)}, respectively. The effect is proportional to the amplitude $n_1$  of the deviation of the density of ions from the bulk value, $\rho_c(z)-\bar\rho_c\approx n_1\exp(-z/\xi_b)$. Figs.~\ref{fig:1} and \ref{fig:2} 
show examples for  the specific choices of the model parameters close to the critical point of the binary solvent,
i.e., for $ \xi_b\gg 1$. The rescaled ratio $R$ between the leading-order correction and the Debye-H\"uckel approximation  is shown in Fig.~\ref{fig:3} for four apirs of the parameters $(\kappa,\xi_b)$.

Let us  consider the non-monotonic correction term  for $\kappa \ll \xi^{-1}_b$, i.e., away from the critical point. 
From (\ref{ra}) it follows that it has a simple exponential form with the amplitude equal to the amplitude  $n_1$:  
\begin{eqnarray}
   \phi^{(2)}(z)/ \phi^{(1)}(z)\approx n_1e^{-z/\xi_b}\hskip 1cm \kappa \ll \xi^{-1}_b.
\end{eqnarray}

 For $T\to T_c$ we have $\xi^{-1}_b\to 0$, and using $\xi^{-1}_b\ll \kappa$ we obtain the approximation
\begin{eqnarray}
\label{eq:anal1}
   \phi^{(2)}(z)/ \phi^{(1)}(z)\simeq \frac{ n_1}{2}\Big[1-\kappa\frac{(1-e^{-z/\xi_b})}{\xi^{-1}_b}\Big]\hskip 1cm \kappa \gg \xi^{-1}_b
\end{eqnarray}
For $z\ll \xi_b$ the above takes the simple form
\begin{eqnarray}
\label{aq:anal2}
   \phi^{(2)}(z)/ \phi^{(1)}(z)\simeq \frac{ n_1}{2}\Big[1-\kappa z\Big]
\end{eqnarray}
The correction term changes sign for  $z_0\approx 1/\kappa$. 
\vspace*{1cm}

Equations for the first corrections to the nonelectrostatic part read
\begin{eqnarray}
\label{cor1}
 \left(\mathbf{v} ^{(2)}\right)^{''}=\mathbf{M}\mathbf{v}^{(2)} +\mathbf{D},
\end{eqnarray}
where $\mathbf{v} ^{(2)}=(\vartheta^{(2)},\eta^{(2)})$, and the components of the vector $\mathbf{D}^T=(D_s,D_{\rho})$ are
\begin{eqnarray}
\label{cor1c}
D_s= \frac{D^a_{\vartheta\vartheta}\vartheta^2+D^a_{\vartheta\eta}\vartheta\eta +D^a_{\eta\eta}\eta^2+D^a_{\phi\phi}\phi^2 }{J}
\end{eqnarray}
\begin{eqnarray}
D_{\rho}= \frac{D^b_{\vartheta\vartheta}\vartheta^2+D^b_{\eta\vartheta}\vartheta\eta +D^b_{\eta\eta}\eta^2+D^b_{\phi\phi}\phi^2  }{J}
\end{eqnarray}

As follows from Eq.~(\ref{cor1})  the correction terms to the composition
profile $\vartheta(z)$ of the mixture are of the order of $e^{-2z/\xi_b }$ and  $e^{-2 \kappa z}$.
Note that the effect of the charges on the  profiles of the composition of the mixture is negligible in the critical region, where $\kappa\gg\xi^{-1}_b$. 

\section{Summary and outlook}
\label{sec:sumout}

Starting from the lattice-gas model of a four-component mixture we have derived the continuum Landau-Ginzburg model for binary mixture solvents in the presence of ions  near the critital point of the demixing transition. The model 
encompasses  the composition of the binary solvent  field $\vartheta$, the density of ions field $\eta$ and the charge density field $\phi$.  It takes into account  electrostatic interactions and the  preferential solvation.
The coupling constants appearing in this extended Landau-Ginzburg theory are given explicitly in terms of thermodynamic quantities, the temperature, the mean composition of solvent,  the mean 
 density of ions, and the interaction parameters $J_{ij}$ characterizing the lattice-gas model of a mixture. We have assumed that ions are of similar chemical nature.
 
The main difference between our functional and the functional studied in Ref.\cite{onuki:04:01} is the presence of the term $\propto \eta^2$ and  terms $\propto (\nabla \eta)^2$ and $ \propto \nabla \eta\nabla \vartheta$ in (\ref{functional}), which result from short-range interaction potentials. These terms lead to the mixing of the fields $\vartheta$ and $\eta$ in the critical order parameter. In the semi-infinite system these terms are important for the form of the profile of the field $\eta$, and through the coupling of $\eta$ with the charge density $\phi$ in the entropy term, they influence the charge profile $\phi(z)$. In our approach, direct couplings between the charge and the concentration are disregarded, but such coupling would be present in the case of ions of different chemical nature (compare Eqs.(\ref{HSR}) and (\ref{HSR1})).

Mean-field theory for our  Landau-Ginzburg model   yields the shift of 
the critical point of the demixing transition with  respect to the case of  binary solvents without ions. The direction of the shift depends on the relative strength of the ions-solvent and solvent-solvent interaction parameters  $J_{s\rho}$ and $J_{ss}$, and is positive for $J_{s\rho}>J_{ss}$.
The linearized EL equations  in the presence of a charged wall do not lead to the effect of the concentration profiles on the charge distribution. We treat nonlinear effects using a perturbation expansion, which gives the simple 
expression for the leading correction to the solution of the linearized EL  Eqs.(\ref{ra}) and (\ref{R}). The ratio between the next-to-leading and leading terms is proportional to the amplitude $n_1$ of the decay of the ion density, $\eta=n_1\exp(- z/\xi_b )$, and otherwise depends only on the dominant lengths in the system, the bulk correlation length $\xi_b$ and the Debye screening length $1/\kappa$.
 We find that 
due to the critical adsorption of that  component of the solvent in which the ions are preferentially dissolved,  the amount of counterions in the layer near the charged surface of the thickness of the Debye screening length
is increased with respect to the one away from the critical point. Critical adsorption of the better solvent  enhances screening  whereas  the critical adsorption of the poorer solvent leads to the opposite effect. This approach is quantitatively valid provided 
the Debye screening length is large,  the  surface-charge density is small, and the surface fields are weak.
We expect the same trend  beyond the approximate perturbation expansion solution. This will be studied further 
by the numerical treatment of the full EL equations.

We should note that our theory shows that the leading-order correction to the Debye-H{\"u}ckel form of the charge profile is determined by the average dielectric constant, i.e., the spatial variation of the dielectric constant can be neglected when $\kappa\ll 1$ (in units of inverse molecular size). 

Finally,  the change of the charge distribution near the charged surface occuring  upon approaching the critical point of the binary solvent, indicates that in  the confining geometries, i.e., for systems between two surfaces,  the effective interactions between the confining surfaces will be altered. 
Our predictions will be tested in the future work by explicit calculations and by comparison with the experimental 
data of Ref.~\cite{bechinger:08:01}.


{\bf Acknowledgments}
We would like to thank S. Dietrich, M. Bier, L. Helden and C. Bechinger for discussions. The work of AC was partially supported by the Polish Ministry of Science and Higher Education, Grant No NN 202 006034.


\end{document}